\documentclass[12pt,a4paper]{article}
\usepackage[T1]{fontenc}        
\usepackage[dvips]{epsfig}
\title{
  {\vspace{-2cm} \normalsize
     \epsfig{figure=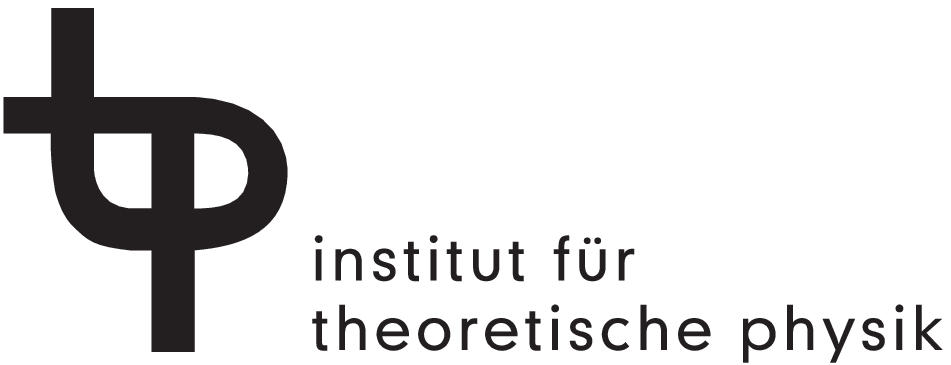,width=80mm}
     \hfill\parbox[b][30mm][t]{35mm}{MS-TP-04-12 \\
                                     hep-lat/0407006}  }\\[25mm]
  On the phase structure of twisted mass lattice QCD
  }
\author{Gernot M\"unster\\
        Institut f\"ur Theoretische Physik,
        Universit\"at M\"unster\\
        Wilhelm-Klemm-Str.~9, D-48149 M\"unster, Germany\\
        e-mail: munsteg@uni-muenster.de}
\date{July 5, 2004}
\newcommand{\I}{\ensuremath{\mathrm{i}\,}}
\newcommand{\E}{\ensuremath{\mathrm{e}\,}}
\begin{document}
\maketitle

\begin{abstract}
The introduction of a chirally twisted mass term has been proposed as an
attractive approach to O(a) improvement of Quantum Chromodynamics with
Wilson fermions on a lattice.  For numerical simulation projects it is
important to know the phase structure of the theory in the region of
small quark masses.  We study this question in the framework of chiral
perturbation theory.  Generalizing the analysis of Sharpe and Singleton
we find extensions of normal and Aoki phase scenarios and a possible new
phase with spontaneous breakdown of chiral symmetry to a discrete
$Z_2$.\\[5mm]
\end{abstract}
\section{Introduction}

In numerical simulations of Quantum Chromodynamics on a lattice one of
the most important issues is to reduce lattice artifacts in order to
improve the approach to the continuum limit.  Recently the lattice
formulation of QCD with Wilson fermions and a chirally twisted quark
mass matrix \cite{TM,Frezzotti} has attracted much interest in this
respect.  Frezzotti and Rossi \cite{FR1,FR2} have advocated this
approach as a way to remove lattice artifacts to first order in the
lattice spacing $a$ by a suitable choice of the chiral twist.
Exploratory numerical studies \cite{QCDtm} have shown favourable
algorithmic features of simulations of twisted mass lattice QCD.

A prerequisite to any numerical simulation project is the knowledge of
the phase structure of the model under consideration.  Where are lines
or points of phase transitions of first or higher order and how do
physical quantities like particle masses behave near them?  It is the
purpose of this note to discuss the phase structure of twisted mass
lattice QCD on the basis of chiral perturbation theory.  Similar
considerations have been made independently by Sharpe and Wu
\cite{ShaWu,Wu}.

We consider lattice QCD with $N_f=2$ quark flavours and degenerate quark
masses $m$.  Let $m_c$ be the ``critical'' value of $m$, where the quark
condensate has a discontinuity, and
\begin{equation}
\tilde{m} = m - m_c
\end{equation}
be the subtracted quark mass.  The chirally twisted mass term can be
introduced in the socalled ``twisted basis'' in the form $\bar{q} (m_c +
M(\omega)) q$ with
\begin{equation}
M(\omega) = m_q \E^{\I \omega \gamma_5 \tau_3}
= \tilde{m} + \I \mu \gamma_5 \tau_3 \,,
\end{equation}
where
\begin{equation}
\tilde{m} = m_q \cos(\omega), \qquad \mu = m_q \sin(\omega)\,,
\end{equation}
and $\tau_a$ denote Pauli matrices.

The low-energy regime of QCD is described by chiral perturbation theory
\cite{Weinberg,GL1,GL2} in terms of the dynamics of the pseudo-Goldstone
bosons of broken chiral symmetry, which are the pions in our case.  The
pion fields $\pi_b(x)$ enter the SU(2)-valued matrices
\begin{equation}
U(x)=\exp \left( \frac{\I}{F_0} \, \pi_b(x) \tau_b \right),
\end{equation}
which are the fundamental variables of chiral perturbation theory.  The
above representation in terms of pion fields refers to the vacuum $U =
{\bf 1}$ and is modified accordingly if the vacuum is at a different
point.  In leading order the chiral effective Lagrangian including
lattice effects up to $\mathcal{O}(a)$ is \cite{Rupak-Shoresh,MS}
\begin{equation}
\mathcal{L}_2 = \frac{F_0^2}{4}\,\mbox{Tr}
\left( \partial_\mu U^\dagger \partial_\mu U \right)
- \frac{F_0^2}{4}\,\mbox{Tr} \left( \chi U^\dagger + U \chi^\dagger \right)
- \frac{F_0^2}{4}\,\mbox{Tr} \left( \rho U^\dagger + U \rho^\dagger \right),
\end{equation}
where
\begin{equation}
\chi= 2 B_0 (\tilde{m} \, {\bf 1} - \I \mu \tau_3),
\end{equation}
appears in the chiral symmetry breaking mass term, and
\begin{equation}
\rho= 2 W_0 a \, {\bf 1}
\end{equation}
parameterizes the lattice artifacts.  The effective Lagrangian in next
to leading order including lattice artifacts up to order $a^2$ has been
calculated by B\"ar, Rupak and Shoresh \cite{BRS}.  The introduction of
a twisted mass term in chiral perturbation and its application to the
calculation of pion masses and decay constants have been made in
\cite{MS,MSS}.

The discussion of the phase diagram of lattice QCD in the framework of
chiral perturbation theory has been pioneered by Sharpe and Singleton
\cite{ShaSi}.  It is based on the potential contained in the chiral
Lagrangian.  Let
\begin{equation}
U = u_0 {\bf 1} + \I u_a \tau_a, \qquad a=1,2,3,
\end{equation}
so that
\begin{equation}
u \equiv (u_0, u_1, u_2, u_3)
\end{equation}
is a unit 4-vector:  $u \cdot u = 1$.  The physical significance of the
variable $u_0$ is given through its relation to the chiral condensate
\begin{equation}
\langle \bar{q} q \rangle = - 2 F_0^2 B_0 \langle u_0 \rangle
\end{equation}
in leading order.
The potential in next to leading order for the case of vanishing twist
is of the form
\begin{equation}
V = -c_1\, u_0 + c_2\, u_0^2\,,
\end{equation}
where
\begin{equation}
c_1 = 2 F_0^2 (B_0 m_q + W_0  a)\,.
\end{equation}
Sharpe and Singleton discuss the region of quark masses, where they are
small of order $a^2$, in which case both coefficients $c_1$ and $c_2$
are to be considered of order $a^2$.

In order to study the phase diagram in the $m_q - \mu$ plane the effect
of the twisted mass term has to be incorporated.  Its addition to the
QCD Lagrangian amounts to extend the potential to
\begin{equation}
V = -c_1\, u_0 + c_2\, u_0^2 + c_3\, u_3\,,
\end{equation}
where
\begin{equation}
c_3 = 2 F_0^2 B_0 \,\mu .
\end{equation}
As for the quark mass, the twist mass $\mu$ is also counted as being of
order $a^2$, so that higher powers of $\mu$ are neglected for the
moment.

Depending on the sign of $c_2$ there are now two possible scenarios for
the phase structure, which will be discussed in the following.

\section{Aoki scenario}

We start by considering the case $c_2 > 0$, where an Aoki phase
\cite{Aoki} appears.  Let us first briefly recall the analysis of Sharpe
and Singleton \cite{ShaSi} for $\mu = 0$.  If
\begin{equation}
| \epsilon | > 1, \qquad \mbox{where} \qquad
\epsilon \equiv \frac{c_1}{2 |c_2|},
\end{equation}
the potential has its minimum outside the allowed region $|u_0| \leq 1$
and consequently the vacuum is at $u_0 = 1, U = {\bf 1}$ or at $u_0 =
-1, U = -{\bf 1}$, depending on the sign of $c_1$.  The pion mass is
given by
\begin{equation}
m_{\pi}^2 = \frac{1}{F_0^2}(|c_1| - 2 c_2).
\end{equation}
Chiral symmetry $\mbox{SU(2)}_L \otimes \mbox{SU(2)}_R$ is broken
explicitly to flavour $\mbox{SU(2)}_V$.

On the other hand, for small quark masses, namely in the range
\begin{equation}
| \epsilon | < 1,
\end{equation}
there is a small region of Aoki phase.  The potential has its minimum at
$u_0 = \epsilon$ and the remaining components of $u$ develop a vacuum
expectation value associated with spontaneous flavour and parity
breaking, which can be chosen to be directed along the third axis:  $u_3
= \pm \sqrt{1 - \epsilon^2}$.  The nonvanishing $u_3$ corresponds to
\begin{equation}
\langle \bar{\chi} \gamma_5 \tau_3 \chi \rangle \neq 0.
\end{equation}
The remaining flavour symmetry is U(1) and $\pi_1$ and $\pi_2$ are the
massless Goldstone bosons.  The neutral pion is massive with a mass
\begin{equation}
m_{\pi 3}^2 = \frac{2 c_2}{F_0^2} (1 - \epsilon^2).
\end{equation}
Varying $c_1$ inside the Aoki phase from $\epsilon = 1$ to $\epsilon =
-1$, the chiral condensate changes continuously from negative to
positive values.

The boundary of the Aoki phase is at $|\epsilon| = 1$, where there is a
second order phase transition and all pions are massless.  The width of
the phase is proportional to $a^3$.

\begin{figure}[hbt]
\vspace{.8cm}
\centering
\epsfig{file=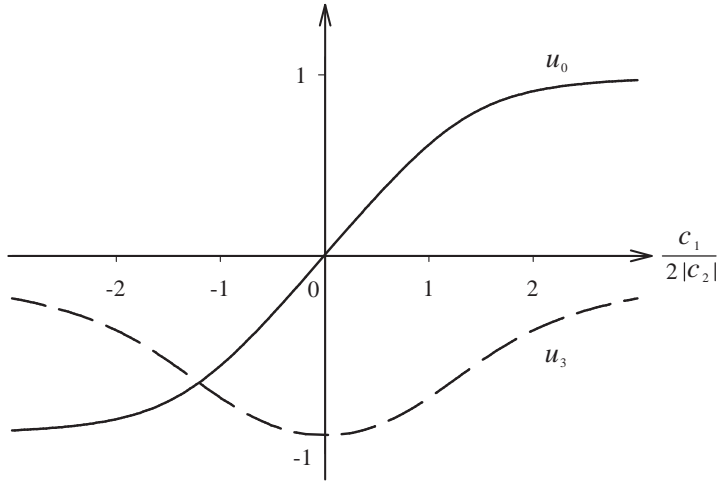,width=10cm}
\caption{\label{u0u3Aoki}
The values of $u_0$ and $u_3$ in the minimum of the potential
as a function of $c_1 / 2 |c_2| \propto \tilde{m}$ for
$c_3 = c_2$ in the Aoki scenario.}
\end{figure}

Now let us switch on the twist mass to a nonvanishing value, which we
may choose to be positive:  $\mu > 0$.  This leads to a shift of the
minimum of the potential in the third direction.  As a consequence, for
all quark masses, $u_3$ acquires a nonvanishing negative value, which is
given by the solution of a quartic equation and is shown in
Fig.~\ref{u0u3Aoki}.

Instead of writing down the exact solution in terms of the formulae of
Ludovico Ferrari \cite{Cardano} here, we restrict ourselves to
displaying it to first order in $c_3 \propto \mu$ for $|\epsilon|$ away
from 1:
\begin{equation}
u_3 = \left\{ \begin{array}{l}
\displaystyle
- \frac{c_3}{|c_1| - 2 c_2} + \mathcal{O}(c_3^2) \qquad
\textrm{for} \quad |\epsilon| > 1,\\
\displaystyle
- \sqrt{1-\epsilon^2} - \frac{\epsilon^2}{1-\epsilon^2}\,
\frac{c_3}{4 c_2} + \mathcal{O}(c_3^2) \qquad
\textrm{for} \quad |\epsilon| < 1.
\end{array} \right.
\end{equation}
The breaking of flavour SU(2) is in this case explicit, and in addition
to $\pi_3$ also the charged pions $\pi_1$ and $\pi_2$ are massive.

\begin{figure}[hbt]
\vspace{.8cm}
\centering
\epsfig{file=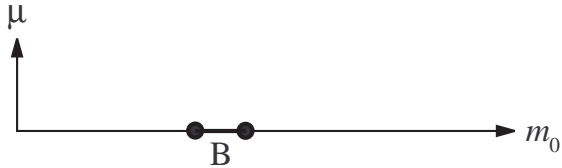,width=8cm}
\caption{\label{Aokiphase}
Phase diagram for the Aoki scenario.}
\end{figure}

$u_0$ changes continuously as a function of the quark mass and there is
no phase transition.  The phase transition, that existed for $\mu=0$, is
washed out for $\mu \neq 0$.  The phase diagram is shown in
Fig.~\ref{Aokiphase}.
Between the second order endpoints is a line of first order transitions,
where $u_3$ makes a jump when going from positive to negative values of
$\mu$.

\section{Normal scenario}

A different scenario results for $c_2 < 0$.  The case of vanishing $\mu$
has been discussed by Sharpe and Singleton \cite{ShaSi} too.  For $\mu =
0$ the potential has its minimum at $u_0 = 1$ for $c_1 > 0$, i.e.\ for
positive quark mass, and at $u_0 = -1$ for $c_1 < 0$.  At $c_1=0$,
corresponding to the quark mass being equal to $m_c$, there is a first
order phase transition, where $u_0$ jumps from $+1$ to $-1$.  The chiral
condensate makes a jump of size $- 4 F_0^2 B_0$.  The pion masses are
nonzero and are given by
\begin{equation}
m_{\pi}^2 = \frac{1}{F_0^2} (|c_1| + 2 |c_2|).
\end{equation}
For $\mu \neq 0$ the minimum of the potential is shifted away from $u_0
= \pm 1$ by the ``magnetic field'' $\mu$.  We have $|u_0| < 1$ and $u_3
\neq 0$.  The corresponding flavour and parity symmetry breaking is
explicit and the pions remain massive.

\begin{figure}[hbt]
\vspace{.8cm}
\centering
\epsfig{file=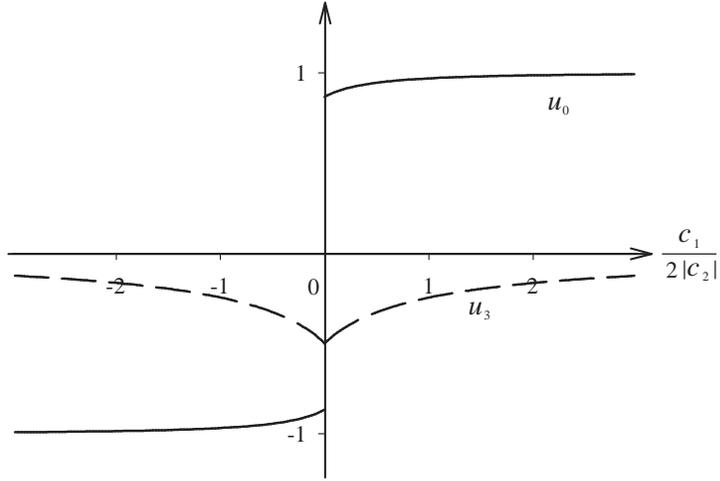,width=10cm}
\caption{\label{u0u3normal}
The values of $u_0$ and $u_3$ in the minimum of the potential
as a function of $c_1 / 2 |c_2| \propto \tilde{m}$ for $c_3 = |c_2|$ in
the normal scenario.}
\end{figure}

The values of $u_0$ and $u_3$ are again given by the solution of a
quartic equation, and are shown in Fig.~\ref{u0u3normal}.

To first order in $\mu$ we have
\begin{eqnarray}
u_3 &=& - \frac{c_3}{|c_1| + 2 |c_2|} + \mathcal{O}(c_3^2)\\
|u_0| &=& 1 - \frac{c_3^2}{2 (|c_1| + 2 |c_2|)^2} + \mathcal{O}(c_3^3).
\end{eqnarray}

\begin{figure}[hbt]
\vspace{.8cm}
\centering
\epsfig{file=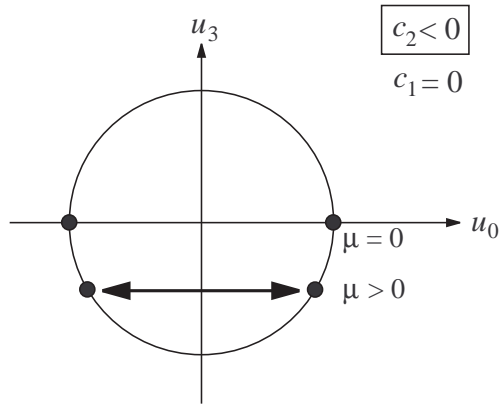,width=7cm}
\caption{\label{u0u3jump}
Jump of $u_0$ at the phase transition for the normal scenario.}
\end{figure}

At $c_1 = 0$ there is a first order phase transition, where $u_0$
changes sign discontinuously, as indicated in Fig.~\ref{u0u3jump}.  The
value of $u_3$ at the transition is
\begin{equation}
u_3 = - \frac{c_3}{2 |c_2|}.
\end{equation}

\begin{figure}[hbt]
\vspace{.8cm}
\centering
\epsfig{file=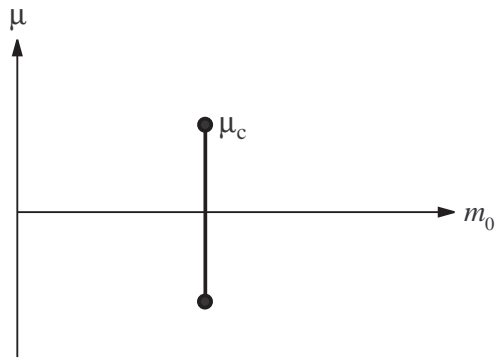,width=7cm}
\caption{\label{normalphase}
Phase diagram for the normal scenario.}
\end{figure}

For increasing $\mu$ the minima of the potential at the transition point
$c_1 = 0$ approach each other.  There is an endpoint of the transition
line at
\begin{equation}
\mu_c = \frac{|c_2|}{F_0^2 B_0} \sim a^2
\end{equation}
and the phase diagram looks like it is shown in Fig.~\ref{normalphase}.

\begin{figure}[hbt]
\vspace{.8cm}
\centering
\epsfig{file=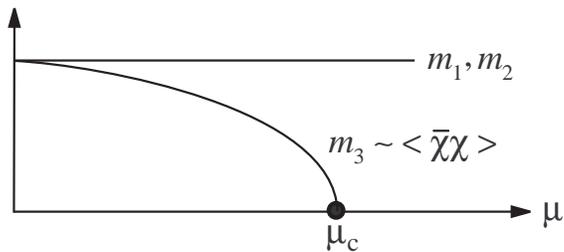,width=8cm}
\caption{\label{mpi-mu}
Pion masses and chiral condensate as a function of $\mu$ along the phase
transition line in the normal scenario.}
\end{figure}

On the phase transition line the neutral pion mass behaves as
\begin{equation}
m_{\pi 3}^2 = \frac{1}{2 F_0^2 |c_2|} ( 4 c_2^2 - c_3^2 ) \sim a^2
\end{equation}
\newpage
and the same holds for the jump of the condensate,
\begin{equation}
\Delta <\bar{\chi} \chi>
= - 4 F_0^2 B_0 \sqrt{1 - \frac{c_3^2}{4 c_2^2}}
= - \frac{4 F_0^3 B_0}{\sqrt{2 |c_2|}} m_{\pi 3}\,,
\end{equation}
as displayed in Fig.~\ref{mpi-mu}.

The masses of the charged pions are continuous across the first order
line, where they have their minimum
\begin{equation}
m_{\pi 1}^2 = m_{\pi 2}^2 = \frac{2 |c_2|}{F_0^2} \sim a^2.
\end{equation}

\section{A new phase}

So far we have considered the case of twist masses $\mu$ of order $a^2$
with only a linear term in $u_3$ in the potential.  In general, higher
terms in the Sharpe-Singleton potential have to be included.  At the
next order additional terms quadratic in $u_a$ have to be added:
\begin{equation}
V = -c_1\, u_0 + c_2\, u_0^2 + c_3\, u_3
+ c_4\, u_3^2 + c_5\, u_0 u_3.
\end{equation}

What are their possible consequences?  In the normal scenario, $c_2 <
0$, one finds that these terms do not lead to qualitative changes of the
phase diagram.  Only numerical modifications will occur.

On the other hand, in the Aoki case the new terms can lead to new
phases, if the coefficient $c_4 \propto \mu^2$ is positive.  In this
situation a minimum of the potential in the interior of the $u_0$--$u_3$
unit circle will occur for sufficiently large $\mu$. As a consequence,
flavour symmetry is broken down spontaneously to a discrete $Z_2$ by
nonvanishing expectation values of $u_1$ or $u_2$. In addition to the
segment of Aoki phase at $\mu = 0$, regions of this new phase in the
upper ($\mu > 0$) or lower ($\mu < 0$) half-planes can form. Inside this
phase the pion masses vanish and the quark condensate varies
continuously.

Along the lines of catastrophe theory \cite{Held}, terms in the
potential of higher orders than quadratic do not change the discussed
scenarios for small masses qualitatively as long as none of the
quadratic coefficients above vanishes accidentally.

\section{Discussion and conclusion}

As we have seen, there can exist a line of first order phase transitions
in the $m_q$--$\mu$ plane, terminating in second order endpoints.
Depending on the sign of the coefficient $c_2$, the line is horizontal
or vertical in the plane.  Whereas for values of the inverse gauge
coupling $\beta < 4.6$ the Aoki scenario appears to be realized
\cite{HU}, recent Monte Carlo calculations \cite{QCDtm} at $\beta =
5.2$ indicate the presence of the normal scenario.  The first order
phase transition line for nonvanishing $\mu$ occurs at a twist angle of
$\pi/2$, where $\mathcal{O}(a)$ improvement is predicted \cite{FR1,FR2}.
The associated two-phase coexistence and metastability hampers numerical
calculations. Nevertheless, interesting physical observables, like
decay constants and squared masses, are continuous across the phase
transition line and are expected to be $\mathcal{O}(a)$ improved there.

From the point of view of Monte Carlo calculations it would be desirable
to have the phase transition line as short as possible.  There should be
a value of $\beta$ somewhere between 4.6 and 5.2 where $\mu_c \propto
c_2$ vanishes.  This point, however, would not be close enough to the
continuum limit.  In order to make $\mu_c$ small one could go nearer to
the continuum limit at larger values of $\beta$, or try to improve the
action in such a way that $c_2$ is small.  One can search for such a
situation on the untwisted axis ($\mu = 0$) by monitoring the minimum of
the pion masses and trying to make it small.  The jump of the condensate
is not suitable for the search, as it is insensitive to $c_2$.
\vspace{8mm}

{\bf Acknowledgement:}
I would like to thank I.~Montvay, F.~Farchioni, S.~Sharpe, J.~Wu and
R.~Frezzotti for stimulating discussions and P.~Hofmann for help with
the figures.


%
\end{document}